\documentclass[11pt,showpacs,preprintnumbers,amsmath,amssymb,prd,nofootinbib,superscriptaddress]{revtex4-2}

\usepackage{dcolumn}
\usepackage{bm}
\usepackage{ifpdf}
\usepackage{hyperref}
\usepackage{bm}
\usepackage{xcolor,color,graphicx,graphics}
\usepackage[spanish,english]{babel}
\usepackage[latin1]{inputenc}
\usepackage[OT1]{fontenc}
\usepackage{latexsym,amssymb,amsmath,amsfonts}
\usepackage{makeidx}
\usepackage{epsfig,subfigure}
\usepackage{natbib}
\usepackage{epstopdf}
\usepackage{mathrsfs}
\usepackage{hyperref}
\hypersetup{colorlinks=true, linkcolor=blue, citecolor=blue, urlcolor=blue}
\usepackage{enumerate}

\everymath{\displaystyle}
\usepackage{graphicx}

\usepackage[T1]{fontenc}
\usepackage{amsmath}
\usepackage{amssymb}
\usepackage{graphicx}
\usepackage{xcolor}
\usepackage{fixmath}
\usepackage{mathtools,slashed}

\newcommand{\bea}{\begin{eqnarray}}
\newcommand{\eea}{\end{eqnarray}}

\newcommand{\orcid}[1]{\href{https://orcid.org/#1}{\includegraphics[width=10pt]{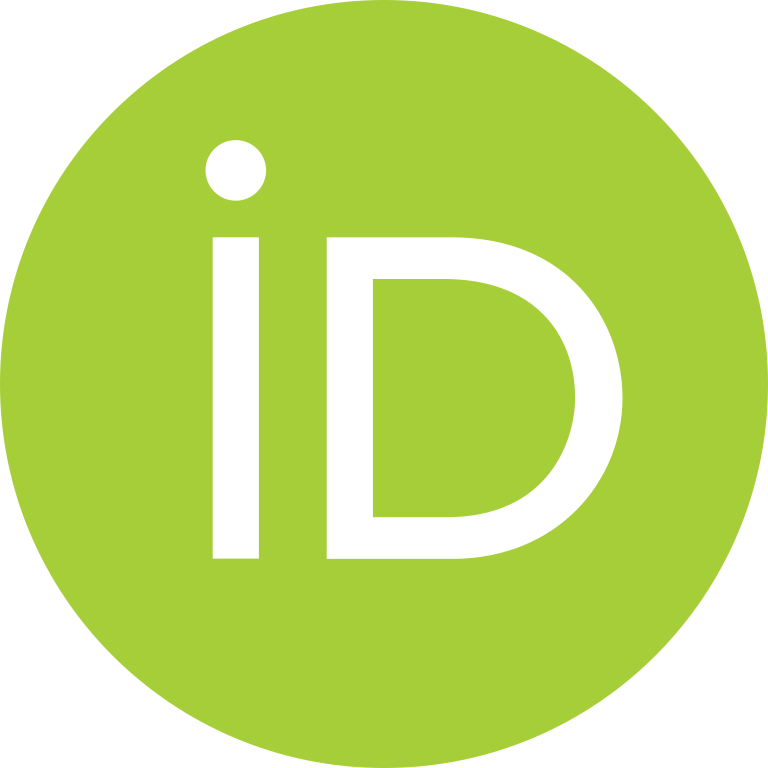}}}

\begin{document}

\title{An attempt to add Barrow entropy in $f(R)$ gravity}

\author{P. S. Ens \orcid{0000-0002-7274-2308}}
\email{peter@fisica.ufmt.br}
\affiliation{Instituto de F\'{\i}sica, Universidade Federal de Mato Grosso,\\78060-900, Cuiab\'{a}, Mato Grosso, Brazil}

\author{A. F. Santos \orcid{0000-0002-2505-5273}}
\email{alesandroferreira@fisica.ufmt.br}
\affiliation{Instituto de F\'{\i}sica, Universidade Federal de Mato Grosso,\\78060-900, Cuiab\'{a}, Mato Grosso, Brazil}

\begin{abstract}	

	In this work, a way to consider together two originally different corrections to the Friedmann equations is presented. The first is the Barrow entropy, which imposes a fractal structure on the black hole horizon area. While the second is the well-known $f(R)$ gravity, which comes from a generalization of the Einstein-Hilbert action. Using the ideas of gravity-thermodynamics conjecture, these two models are combined. Then the modified Friedmann equation is obtained. Choosing a particular $f(R)$ model, an application is investigated. The state parameter and the density parameters for matter and dark energy are calculated. With these results,  the dynamic evolution of the universe is discussed.
	 
\end{abstract}

\maketitle

\section{Introduction}

Gravity is one of the fundamental forces in the universe. However, there is no complete theory applicable to all energy regimes that describes this interaction. Classically, the theory of general relativity is the best gravitational theory that responds very well to various observational tests \cite{Will}. Although attempts at a quantum version have been built and discussed for a long time, a completely consistent theory of quantum gravity has not been obtained. At low energies, some effective theories have been investigated \cite{Don, Don1, Don2, Don3, Bur}. Another study that plays an important role in providing fundamental information about the quantum aspect of gravity is the black hole thermodynamics. Early works by Hawking and Bekenstein showed that a black hole radiates at a given temperature and that its entropy is proportional to the area of the black hole's event horizon \cite{Bek1, PhysRevD.7.2333, Haw}. This leads to the possibility that  the black hole thermodynamics can be investigated in the context of quantum gravity. Furthermore, since the beginning of discussions about possible relations between general relativity and thermodynamics, ideas have emerged on how to obtain equations to better describe the different periods throughout the expansion of the universe. One of them is the holographic principle, which consists of studying the behavior of a system based on the characteristics of its surface \cite{RevModPhys.74.825}. Theories like this aim to change the way entropy is calculated \cite{PhysRevD.7.2333, PhysRevD.13.191}. From these discoveries, other speculations have been analyzed, such as the relationship between thermodynamics and Einstein equations, which is known as the gravity-thermodynamics conjecture.

The gravity-thermodynamics conjecture is a formalism that allows obtaining Einstein equations from a thermodynamic approach, using the relationship between entropy and the area of the event horizon of a black hole \cite{cai, Jac, Pad1, Pad2}. Taking the first law of thermodynamics on the apparent horizon, the Friedmann equations can be derived. This procedure is applicable in general relativity as well as in a variety of modified gravity theories. However, when the analysis is extended to a modified theory, the entropy relation usually changes \cite{Wang, Akbar, faraoni2, wald, oneloop}. Since the Bekenstein-Hawking entropy-area law is a non-extensive measure, various generalized statistical mechanics have been proposed to study cosmic evolution and, in general, the gravitational phenomena. Inspired by this, some models have been developed, such as Tsallis entropy \cite{Tsallis1988, Tsallis2013}, R\'{e}nyi entropy \cite{Renyi}, among others. Recently, a different proposal, known as Barrow entropy \cite{barrow0}, has been considered. In this work, Barrow entropy and its consequences for the evolution of the universe will be studied.

In order to build his model, Barrow was inspired by the geometrical structure of the COVID-19 virus \cite{barrow0}. In this model, quantum-gravitational effects can introduce intricate fractal features into the black-hole structure. This structure leads to a fractal horizon surface. Then a new black hole entropy relation is defined, i.e. $S\sim A^{f(\Delta)}$, where $f(\Delta)=1+\Delta/2$. The $\Delta$ parameter quantifies the quantum-gravitational deformation. In this context, some applications have been investigated. For example, Baryon asymmetry has been studied \cite{Luciano:2022pzg}, inflation driven by Barrow holographic dark energy has been considered \cite{Maity:2022gdy}, early and late periods of the universe from a new generalized entropy have been analyzed \cite{NOJIRI2022137189}, Barrow holographic dark energy has been formulated \cite{Sari}, the cosmology using Barrow entropy has been proposed \cite{Leon_2021}, the generalized second law of thermodynamics with Barrow entropy has been investigated \cite{Basi}, among others. In this context, the main objective of this paper is to analyze the dynamic evolution of the universe considering the Barrow entropy in $f(R)$ gravity. 

The motivations for considering an alternative gravity theory to general relativity arise from various observational data \cite{Per, Riess, Wmap, Blake} that lead to the accelerated expansion of the current universe. One of the simplest and most studied models of modified gravity is the $f(R)$ theory \cite{faraoni1}. In this modified theory, the gravitational action of general relativity takes a general form by incorporating an arbitrary function of the Ricci scalar into it. This is not the only modified theory of gravity, there are several theories in the literature that generalize the general relativity, such as $f(R,T)$ gravity \cite{PhysRevD.84.024020}, gravity theories with non-minimally coupled scalar fields \cite{Sotiriou2015}, among many others. Although there are many theories of gravity, our interest here is to study $f(R)$ gravity combined with Barrow entropy using the ideas of the gravity-thermodynamics conjecture and then describe the evolution of the universe from there.

This paper is organized as follows. In section II, Barrow entropy and its equations are introduced. Using the gravity-thermodynamics conjecture, the Friedmann equation is derived for two different situations. First, the Hawking-Bekenstein entropy is used. And then, consider Barrow entropy. In section III, $f(R)$ theory is presented. The first law of thermodynamics is discussed. In section IV, $f(R)$ theory and Barrow entropy are put together. It is assumed that the energy density and pressure of dark energy are composed of two parts, one due to the modified gravity and the other due to Barrow entropy. In this context, a modified Friedmann equation is obtained. In section V, an application is investigated considering the generalized Carroll-Duvvuri-Trodden-Turner model. Then the state parameter is calculated. In section VI, some concluding remarks are made.

\section{Barrow entropy and its equations}

In this section, Friedmann equations are obtained using the gravity-thermodynamics conjecture \cite{akbar0}. Starting from the first law of thermodynamics, two different cases are investigated. In the first case, the Bekenstein-Hawking entropy is used, while in the second case the Barrow entropy is considered. The first law of thermodynamics is given as
	\begin{equation}
		TdS=dE-WdV ,
	\end{equation}
	where $T$ is temperature, $S$ is entropy, $E$ is the  energy, $V$ is the volume and  $W=(\rho-p)/2$ is the work, with $\rho$ and $p$ being the energy density and pressure, respectively. To apply the first law of thermodynamics, temperature and entropy need to be known. The standard choice
is to use the Bekenstein-Hawking entropy, i.e. $S=\frac{A}{4G}$, with $A=4\pi r^2$ the area of the black hole horizon and $G$ the gravitational constant \cite{ross}. 
For the temperature is chosen $T=-\frac{1}{2\pi r}\left(1-\frac{\dot{r}}{2Hr}\right)$, the horizon temperature  \cite{cai,Helou}. Here $r$, the apparent horizon, would generally be $r=\frac{1}{\sqrt{H^2+k/a^2}}$, but since we are dealing with the flat case of FRW metric, implying $k=0$, we have $Hr=1$. Thus the temperature can be written in a simpler form $T=-\frac{1}{2\pi r}\left(1-\frac{\dot{r}}{2}\right)$. Then the first law of thermodynamics becomes
	\begin{equation}\label{eqterm0}
		-\frac{1}{2\pi r} \left(1-\frac{\dot{r}}{2}\right) \frac{dA}{4G}=dE-\frac{\rho-p}{2}dV .
	\end{equation}
Imposing the matter conservation equation, i.e.  $\dot{\rho}+3H(\rho+p)=0$, Eq. (\ref{eqterm0}) is written as
\begin{equation}
		-\frac{1}{2\pi r} \left(1-\frac{\dot{r}}{2}\right) \frac{dA}{4G}=-A(\rho+p)dt+\frac{\rho+p}{2}dV ,
	\end{equation}
	where $dE=Vd\rho + \rho dV = -A(\rho+p)dt+\rho dV$ has been used. Using $dV=Adr=A\dot{r}dt$ leads to
	\begin{equation}
		\frac{\dot{r}}{r^2}= 4\pi G(\rho+p).\label{pre}
	\end{equation}
	In a flat universe, $\frac{\dot{r}}{r^2}=-\dot{H}$, then Eq. (\ref{pre}) becomes
	\begin{equation}
		\dot{H}= -4\pi G(\rho+p).
	\end{equation}
	This is the Friedmann equation. Applying it to the conservation relation and performing the integration in time, results in
	\begin{equation}
		H^2 = \frac{8\pi G}{3} \rho + C ,
	\end{equation}
	where $C$ acts as the cosmological constant.
	
	It is important to note that, considering a fixed boundary ($\dot{r}=0$) with an energy flux through it \cite{cai}, there is another way to obtain the same Friedmann equation. In this case, the temperature is defined as $T=\frac{1}{2\pi r_A}$, for a static system, and the first law of thermodynamics is used as $dE = TdS$.
	
	There are some proposals for obtaining new  equations of motion. One of them is to consider modifications in the black hole area, as proposed by Barrow \cite{barrow0}. In this model, the area takes into account a possible fractal structure from the surface roughness, which depends on a parameter $\Delta$ representing its intricacy. This leads to a new form for entropy which is given as
	\begin{equation}
		S=\left(\frac{A}{4G}\right)^{1+\Delta/2} .
	\end{equation}
	
From the Barrow entropy, the equation of motion or the modified Friedmann equation can be obtained following the same procedure applied previously. Let us consider the entropy defined in a more general way, i.e. $S=\frac{f(A)}{4G}$. In our case,  $f(A)=A^{1+\frac{\Delta}{2}}/(4G)^{\frac{\Delta}{2}}$ is chosen to match Barrow definition. Starting from the first law of thermodynamics it is found
	\begin{equation}
		-\frac{1}{2\pi r} \left(1-\frac{\dot{r}}{2}\right) \frac{df(A)}{4G}=dE-\frac{\rho-p}{2}dV .
	\end{equation}
	Since $df(A)=f'(A)dA$, the last equation reads
	\begin{equation}
		-f'(A)\frac{1}{2\pi r} \left(1-\frac{\dot{r}}{2}\right) \frac{dA}{4G}=dE-\frac{\rho-p}{2}dV .
	\end{equation}
	It should be noted that this equation is similar to Eq.  (\ref{eqterm0}). Then, following the same procedure as before, it is obtained that
	\begin{equation}\label{eqcampofAgen}
		f'(A)\dot{H} = -4\pi G (\rho + p) .
	\end{equation}
	This is the modified Friedmann equation due to Barrow entropy. To obtain the second Friedmann equation,  Eq.  (\ref{eqcampofAgen}) is applied in the energy conservation relation. Then
	\begin{equation}
		\dot{\rho}=\frac{3}{4\pi G}f'(A)H\dot{H} .
	\end{equation}
	Performing the integration in time leads to
	\begin{equation}\label{eqcampofAgen2}
		\frac{2}{2-\Delta}f'(A)H^{2} = \frac{8\pi G}{3}\rho + \frac{\Lambda}{3} ,
	\end{equation}
	where  $\Lambda=3\frac{2}{2-\Delta}f'(A_0)H_0^{2}-8\pi G \rho_0$ is the integration constant. Here $\rho_0$ and $H_0$ are the energy density of the fluid and the Hubble parameter, respectively, defined as initial conditions of the system, typically at present time, furthermore, $A_0=4\pi H^{-2}_0$ is the horizon area calculated from these conditions. It is important to note that this form of Eq. \eqref{eqcampofAgen2} depends exclusively on the function $f(A)$.
	
	These equations can be written in the same structure as the usual Friedmann equations evidencing the presence of dark energy originating from the new entropy. Then Eq. (\ref{eqcampofAgen}) is given as
	\begin{equation}
		\dot{H} = -4\pi G \left(\rho + p - \frac{\dot{H}}{4\pi G}(1-f'(A))\right) ,
	\end{equation}
	and Eq. (\ref{eqcampofAgen2}) becomes
	\begin{equation}\label{eqcampofAgen3}
		H^2 = \frac{8\pi G}{3}\left(\rho + \frac{\Lambda}{8\pi G} + \frac{3H^2}{8\pi G}\left(1-\frac{2}{2-\Delta}f'(A)\right)\right) .
	\end{equation}
	As a result, the components of dark energy are 
	\begin{equation}\label{rhoDE_barrow}
		\rho_{DE} = \frac{1}{8\pi G}\left(\Lambda(\Delta) + 3H^2 \left(1 - \frac{2}{2-\Delta}f'(A)\right)\right) ,
	\end{equation}
	\begin{equation}\label{PDE_barrow}
		p_{DE} = - \frac{1}{8\pi G} \biggl(\Lambda(\Delta) + 2\dot{H} \left(1 - f'(A)\right) + 3H^2 \left(1 - \frac{2}{2-\Delta}f'(A)\right)\biggr) ,
	\end{equation}
	with $\rho_{DE}$ and $p_{DE}$ being the Barrow's holographic dark energy density and pressure, respectively \cite{barrow1}. It should be emphasized that for Eq. (\ref{eqcampofAgen3}) as for $\rho_{DE}$ and $p_{DE}$, it is considered specifically $f(A)=A^{1+\frac{\Delta}{2}}/(4G)^{\frac{\Delta}{2}}$.
	
	It is important to note that, although Eqs. (10)-(12) are enough to study the dynamics of the universe, the manipulations of Eqs. (13)-(14) that lead to the definitions of Eqs. (15)-(16) are necessary to have the conservation relations that are given as $\dot{\rho}_m+3H(\rho_m+p_m)=0$ and $\dot{\rho}_{DE}+3H(\rho_{DE}+p_{DE})=0$. Consequently, also making it clear that these two sectors are not interacting, a similar result is shown in references \cite{Paul,Sari2,Sharma,Fara}.
	
	As in the Bekenstein-Hawking case, a static system with an energy flux crossing its boundary can be considered, as in \cite{barrow1}. Assuming $T=\frac{1}{2\pi r_A}$ and the relation $-dE = TdS$, it is possible to arrive at the same equation of motion Eq. (\ref{eqcampofAgen3}) and dark energy components Eqs. (\ref{rhoDE_barrow}) and (\ref{PDE_barrow}).
	
	In the next section, the $f(R)$ gravity in the context of the  gravity-thermodynamics conjecture is investigated.

\section{$f(R)$ gravity and the first law of thermodynamics}

Here the $f(R)$ gravity is considered. Then the effective energy density and pressure are obtained. Using these quantities, the first law of thermodynamics is investigated. In such a discussion, a correction for the black hole entropy due to $f(R)$ gravity is analyzed. 

Let's start with the action that describes this modified gravitational theory. This model ge\-neralizes the Einstein-Hilbert action, replacing the Ricci scalar $R$ by a function of the Ricci scalar $f(R)$ , i.e.,
	\begin{equation}
		S=\int f(R)\sqrt{-g}d^4x .
	\end{equation}
	Field equations are  derived from the variational principle  \cite{faraoni1}.  Varying the action with respect to the metric, the field equations are obtained as
	\begin{equation}\label{eqcampofR}
		R_{\mu\nu}f'(R)-\frac{1}{2}g_{\mu\nu}f(R)-\left(\nabla_\mu\nabla_\nu+g_{\mu\nu}\Box \right)f'(R)= -\kappa T_{\mu\nu},
	\end{equation}
	where $\kappa=8\pi G$.
	Taking a perfect fluid as matter content and a flat FRW universe, the field equations are given as
	\begin{equation}
		H^2 = \frac{\kappa}{3f'(R)} \left(\rho_{M} + \rho_{eff}\right) ,\label{eq1}
	\end{equation}
	\begin{equation}
		2\dot{H} + 3H^2 = -\frac{\kappa}{f'(R)} \left(p_{M} + p_{eff}\right) ,\label{eq2}
	\end{equation}
where  the effective density and pressure are defined, respectively, as
	\begin{equation}
		\rho_{eff} = \frac{1}{\kappa} \left(\frac{Rf'(R)-f(R)}{2} - 3H\dot{R}f''(R)\right) ,
	\end{equation}
	\begin{equation}
		p_{eff} = \frac{1}{\kappa} \left(\dot{R}^2f'''(R) + \ddot{R}f''(R) - \frac{Rf'(R)-f(R)}{2} + 2H\dot{R}f''(R)\right) .
	\end{equation}
These definitions of $\rho_{eff}$ and $p_{eff}$, after some calculation considering $\dot{f}'(R)=\dot{R}f''(R)$ and that $\frac{R}{2}-3\dot{H}=6H^2$, lead to a non-conservation energy relation given by
	\begin{equation}\label{conservacaofR1}
		\dot{\rho}_{eff}+3H(\rho_{eff}+p_{eff})=\frac{3}{\kappa}H^2\dot{f}'(R) .
	\end{equation}
	
	Substituting Eq. (\ref{eq1}) in Eq. (\ref{eq2}) leads to
	\begin{equation}\label{friedmannrhopfR}
		dH = -\frac{\kappa}{2f'(R)} \left(\rho_{M} + p_{M} + \rho_{eff} + p_{eff}\right)dt.
	\end{equation}
	Considering the relation $Hr=1$, the last equation is written as
	\begin{equation}
		\frac{f'(R)}{G}dr = A \left(\rho_{M} + p_{M} + \rho_{eff} + p_{eff}\right)dt .\label{25}
	\end{equation}
	
	In this gravitational theory, the black hole entropy and its relation to the horizon area is corrected \cite{faraoni2,wald,oneloop}. In this context, the entropy is given as
	\begin{equation}
	S = \frac{Af'(R)}{4G}.\label{newS}
	\end{equation}
	Using this modified black hole entropy, Eq. (\ref{25}) becomes
	\begin{equation}\label{semdeftemp}
		\frac{1}{2\pi r}dS - \frac{1}{2\pi r}\frac{A}{4G}df'(R) = A \left(\rho_{M} + p_{M} + \rho_{eff} + p_{eff}\right)dt .
	\end{equation}
	
	Now, taking the horizon temperature, i.e. $T=-\frac{1}{2\pi r}\left(1-\frac{\dot{r}}{2}\right)$, the thermodynamics relation (\ref{semdeftemp}) reads
	\begin{equation}\label{reltempunruh1}
		TdS - T\frac{A}{4G}df'(R) = -A \left(\rho+p\right)dt + \frac{A}{2} \left(\rho+p\right)dr .
	\end{equation}
	For convenience, it is defined $\rho=\rho_{M} + \rho_{eff}$ and $p=p_{M} + p_{eff}$. 
	
	Adding the non-conservation relation for $f(R)$ gravity Eq. (\ref{conservacaofR1}) to that of matter, i.e. $\dot{\rho}_M+3H(\rho_{M}+p_M)=0$, we get
	\begin{equation}
		d\rho = - 3H(\rho+p)dt + \frac{3}{8\pi G}H^2\dot{f}'(R)dt .
	\end{equation}
	Thus, the differential of the energy, which is given as $dE=\rho dV + Vd\rho$, results in
	\begin{equation}
		dE= \rho Adr - A(\rho+p)dt + \frac{1}{2\pi r}\frac{A}{4G}df'(R) .\label{dE}
	\end{equation}
	Using Eq. (\ref{dE}) in Eq. (\ref{reltempunruh1}) we obtain
	\begin{equation}
		TdS = dE - WdV + T\left(\frac{4-\dot{r}}{2-\dot{r}}\right)\frac{A}{4G}df'(R).
	\end{equation}
	This is the first law of thermodynamics generated in the context of $f(R)$ gravity. However, it is important to note that, an extra term $Td\bar{S}$ with $d\bar{S}=-\left(\frac{4-\dot{r}}{2-\dot{r}}\right)\frac{A}{4G}df'(R)$ arises. There are some discussions in the literature \cite{felice,akbar} which argue that this extra term may originate from the entropy generated internally by the out-of-equilibrium system.

	As an attempt to obtain an energy conservation relation in $f(R)$ theory and determine the first law of thermodynamics without an extra term that comes from an out-of-equilibrium system, let us write the modified Friedmann equations, given in Eqs. (\ref{eq1}) and (\ref{eq2}),  following Barrow definition. Then these equations become
	\begin{equation}
		H^2 = \frac{\kappa}{3} \left(\rho_{M} + \rho_{DE}\right) ,
	\end{equation}
	\begin{equation}
		2\dot{H} + 3H^2 = -\kappa \left(p_{M} + p_{DE}\right) ,
	\end{equation}
	where
	\begin{equation}\label{rhoDE_f(R)}
		\rho_{DE} = \frac{1}{\kappa} \left(\rho_{eff} + 3H^2(1-f'(R))\right) ,
	\end{equation}
	\begin{equation}\label{PDE_f(R)}
		p_{DE} = \frac{1}{\kappa} \left(p_{eff} + 2\dot{H}(1-f'(R)) + 3H^2(1-f'(R))\right) .
	\end{equation}
	It should be noted that Eqs. (\ref{rhoDE_f(R)}) and (\ref{PDE_f(R)}) describe holographic dark energy due to the $f(R)$ theory.
	These definitions lead to $\dot{\rho}_{DE}+3H(\rho_{DE}+p_{DE})=0$. Performing the same procedure as before, one finds
	\begin{equation}\label{primeiralei}
		TdS = dE - WdV .
	\end{equation}
	Therefore, the usual first law of thermodynamics is obtained. Furthermore, the entropy correction due to the $f(R)$ theory, Eq. (\ref{newS}), is not used.
	
	In the next section, Barrow entropy is taken to $f(R)$ gravity. In this context, the dynamics of the universe is investigated.

\section{Barrow entropy and $f(R)$ theory}

	Barrow's proposal is elegant and carries good motivations. However, there are obstacles that prevent direct application to explain the current behavior of the universe and its history. One of them is the fact that in all cases this modification always leads to a de-Sitter expansion \cite{Leon_2021}. Another problem is the drastic restriction on the possible values of $\Delta$, when considering the characteristics of the Big Bang \cite{barrow1, universe8020102}, as well as other observations,  such as those given in  \cite{Vagnozzi:2022mo}. Likewise, there are also several problems present in the $f(R)$ gravity models, for example, general viability conditions such as $f'(R)>0$ and $f''(R)>0$ to avoid ghost scalar fields and complex mass values for the scalaron field.  For a review of $f(R)$ gravity and its problems see  \cite{felice}.
		
	As both proposals present some problems, the main objective of this section is to propose a way to merge these models, in an attempt to obtain results that are not subject to the problems each one carries individually.
	
	Looking at the definitions given in Eqs.  (\ref{rhoDE_barrow}), (\ref{PDE_barrow}), (\ref{rhoDE_f(R)}) and (\ref{PDE_f(R)}), it is possible to perceive similarity in how the terms originated from their respective modifications act as dark energy. Taking this into account, a new definition for holographic dark energy, which carries the Barrow and $f(R)$ modifications, is considered, that is,
	\begin{equation}
		\rho_{DE} =\frac{1}{\kappa}\left(\rho_{eff} + 3H^2\left(1-\frac{2}{2-\Delta}f'(A)f'(R)\right)\right) ,\label{nrho}
	\end{equation}
	\begin{equation}
		p_{DE} = \frac{1}{\kappa}\biggl(p_{eff} - 2\dot{H}\left(1-f'(A)f'(R)\right) - 3H^2\left(1-\frac{2}{2-\Delta}f'(A)f'(R)\right)\biggr) ,\label{np}
	\end{equation}
	where Eqs. (\ref{nrho})  and (\ref{np}) satisfy the relation $\dot{\rho}_{DE}+3H(\rho_{DE}+p_{DE})=0$. In order to obtain the modified Friedmann equations, let us start with the first law
	\begin{equation}
		TdS = dE - WdV ,
	\end{equation}
	with $dE=\rho dV+Vd\rho$ and $W=(\rho-p)/2$. Then
	\begin{equation}
		TdS = Vd\rho + \frac{\rho+p}{2}dV .
	\end{equation}
	Assuming $\rho=\rho_M+\rho_{DE}$ and $p=p_M+p_{DE}$ and using the energy conservation relation, the last equation becomes
	\begin{equation}
		TdS = -\left(1-\frac{\dot{r}}{2}\right)A(\rho+p)dt .
	\end{equation}
	Taking the horizon temperature $T=-\frac{1}{2\pi r}\left(1-\frac{\dot{r}}{2}\right)$ and entropy $S=A/4G$ we have
	\begin{equation}
		-\frac{1}{2\pi r}\left(1-\frac{\dot{r}}{2}\right)\frac{dA}{4G} = -\left(1-\frac{\dot{r}}{2}\right)A(\rho+p)dt.
	\end{equation}
	After some steps, it is written as
	\begin{equation}
		\frac{\dot{r}}{r^2} = 4\pi G(\rho+p) .
	\end{equation}
	Using $r=H^{-1}$ and $\dot{r}=-H^{-2}\dot{H}$ lead to 
	\begin{equation}
		\dot{H} = -4\pi G(\rho+p) ,\label{Feq}
	\end{equation}
	that by bringing the definitions of densities and pressures, Eq. (\ref{Feq}) is rewritten as
	\begin{equation}\label{friedmannrhopfRbarrow}
		f'(A)\dot{H} = -\frac{\kappa}{2f'(R)}\left(\rho_{M} + p_{M} + \rho_{eff} + p_{eff}\right) .
	\end{equation}
	Applying to the conservation equation and performing a temporal integration results in
	\begin{equation}\label{friedmannrhofRbarrow}
		\frac{2}{2-\Delta}f'(A)H^2 = \frac{\kappa}{3f'(R)}\left(\rho_{M} + \rho_{eff}\right) .
	\end{equation}
	
	It should be noted that  Eqs.  \eqref{friedmannrhopfRbarrow} and \eqref{friedmannrhofRbarrow} are reduced to the usual form for both Barrow and $f(R)$ individually when considering $\Delta=0$ or $f(R)=R$, respectively.
	
	Another analysis is possible. If the non-equilibrium case is desired, the non-conservation relation is used, i.e.
	\begin{equation}
		\dot{\rho}_{eff} + 3H(\rho_{eff}+p_{eff}) = \frac{3}{8\pi G}\frac{2}{2-\Delta}f'(A)H^2\dot{f}'(R).
	\end{equation}
	In addition, entropy correction 
	\begin{equation}
		S = \frac{f(A)}{4G}f'(R) 
	\end{equation}
	is considered. This definition for entropy comes from the fact that while the Barrow correction acts on how the horizon area is calculated \cite{barrow0}, the $f(R)$ correction acts on the physics of the system, specifically as a correction of the gravitational constant \cite{faraoni1}, as $G_{eff}=G/f'(R)$. Furthermore, this form of entropy also fits the proposal of dark energy components to result in a thermodynamic relation that recovers all the characteristics of the Barrow and $f(R)$ cases individually.  With these ingredients, the thermodynamic relation reads
	\begin{equation}
		TdS = dE - WdV + T\left(\frac{4-\dot{r}+\Delta\dot{r}/2}{2-\dot{r}-\Delta\left(1-\dot{r}/2\right)} \right)\frac{f(A)}{4G}df'(R) .
	\end{equation}
	As discussed earlier, the last term can originate from an out-of-equilibrium system \cite{felice,akbar}.

As an application, in the next section, the Barrow-$f(R)$ field equation will be solved and the state parameter in this context will be analyzed.

\section{Application on the generalized Carroll-Duvvuri-Trodden-Turner model}	
	
	To get an idea of how the addition of Barrow entropy can influence an $f(R)$ model, Eq. (\ref{friedmannrhofRbarrow}) is numerically solved using the generalized Carroll-Duvvuri-Trodden-Turner (CDTT) model \cite{PhysRevD.70.043528, PhysRevD.71.063513, Ens_2020}. The $f(R)$ function that describes this model is defined as
	\begin{equation}
		f(R) = R + \lambda R^2 + \frac{\mu}{R},
	\end{equation}
	where $\lambda$ is a constant and $\mu$ is a parameter with units of mass. This model is chosen because it presents two distinct expansion periods. Initially dominated by the term $\lambda R^2$, which causes an exponential expansion. Later, it is dominated by $\mu/R$, which leads to asymptotic behavior with a power-law expansion.
	
\begin{figure}[tb]
\begin{center}
	\includegraphics[scale=0.25]{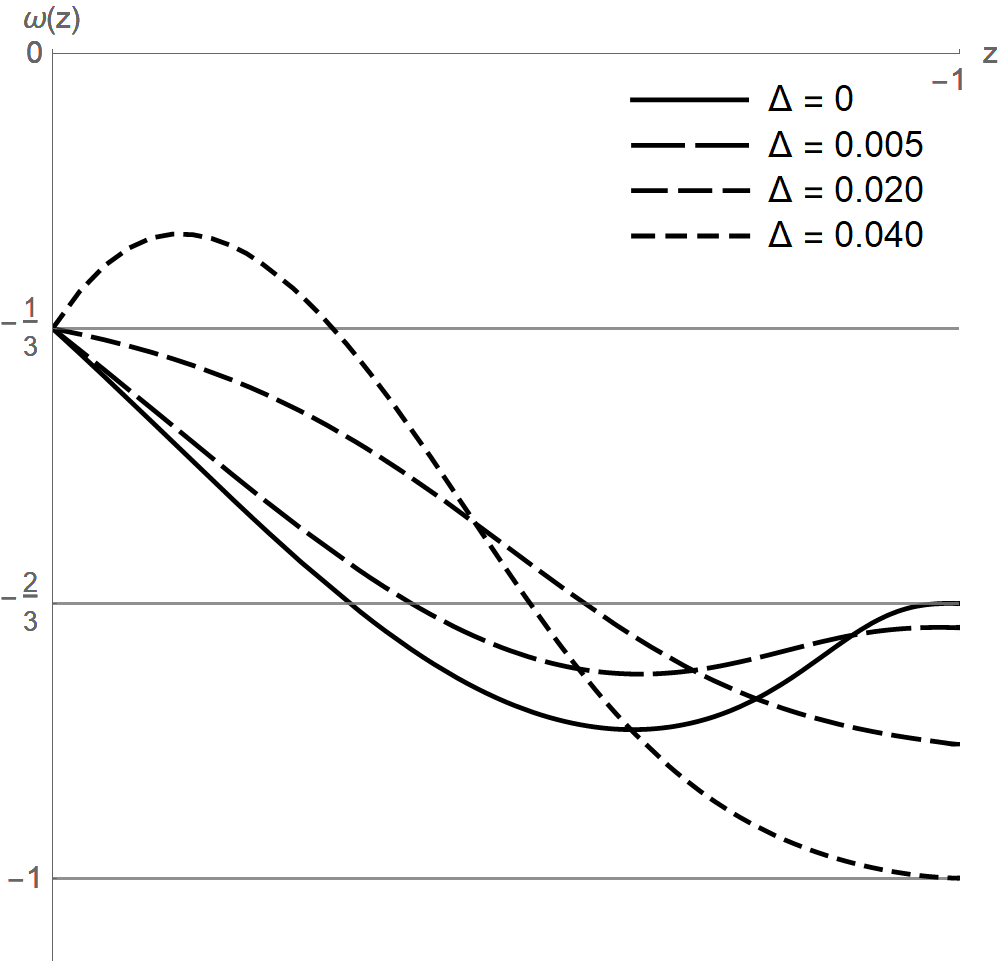} \quad\quad\quad
	\includegraphics[scale=0.25]{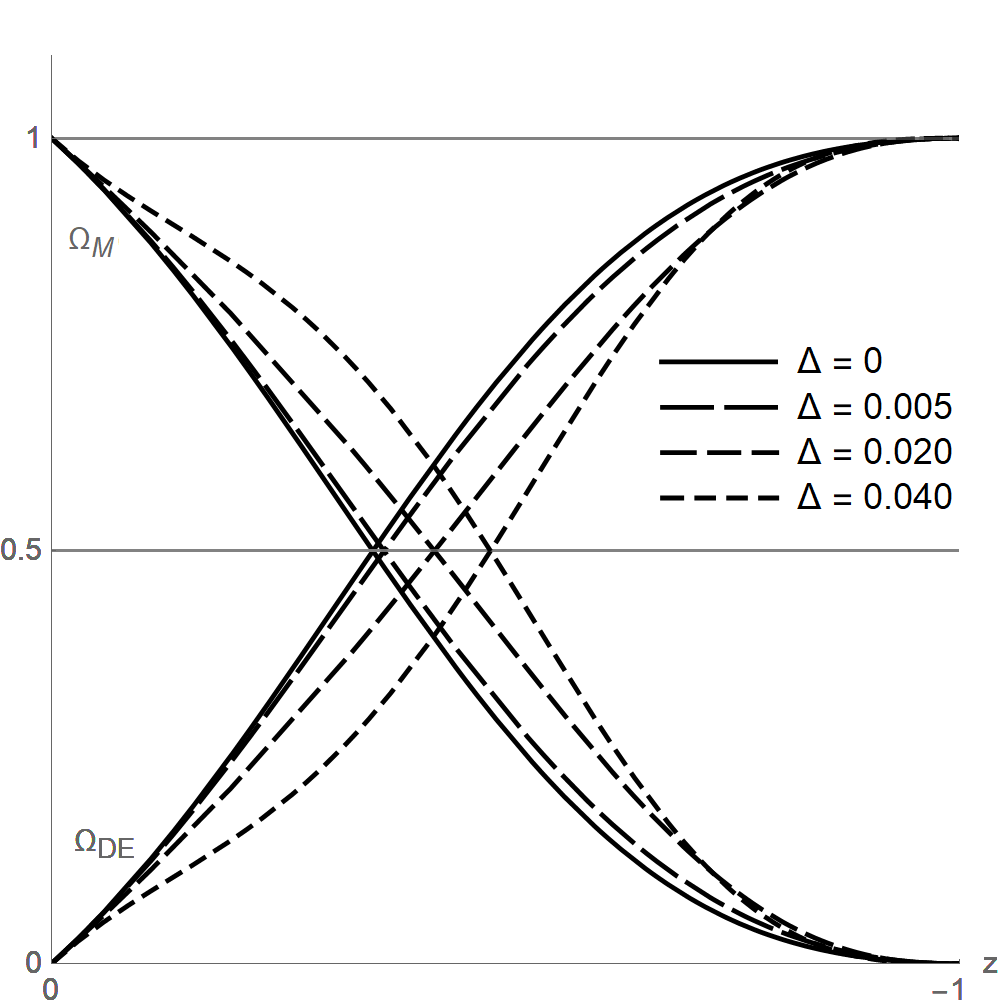}
\caption{\label{figura1}Left side: State parameter $\omega$ versus redshift $z$. Right side:  Matter ($\Omega_M$) and dark energy ($\Omega_{DE}$) densities versus redshift. Here it is considered gCDTT + Barrow models with $H_0=0.1$, $\lambda=3\times10^3$, $\mu=9.7\times10^{-2}$, $\Omega_{M}(0)=1$, $\Lambda=0$.} 
\end{center}
\end{figure}

	Solving the field equation, i.e. Eq.(\ref{friedmannrhofRbarrow}), the scale factor is obtained, which is then used to calculate the state parameter, that is defined as
	\begin{equation}
		\omega = \frac{p_M + p_{DE}}{\rho_M + \rho_{DE}}.
	\end{equation}
	In addition, the density parameters,
	\begin{equation}
		\Omega_M = \frac{8\pi G}{3H^2}\rho_M \quad\quad  \mathrm{and} \quad\quad \Omega_{DE} = \frac{8\pi G}{3H^2}\rho_{DE},
	\end{equation}
	 for matter and dark energy, respectively, are calculated. In Figure \ref{figura1} the evolution of the state and density parameters are compared using different values of $\Delta$.
	
	The initial conditions are defined so that the system always starts from an inertial expansion, $\omega=-1/3$, to better visualize the influence of the dark energy. By default, this $f(R)$ model produces an initial acceleration that approaches to exponential, $\omega=-1$, depending on the choosen values of $\lambda$, and eventually it stabilizes itself to a power law, $\omega=-2/3$, according to the parameter $\mu$. All this can be observed by the behaviour of curve $\Delta=0$. Note that the presence of the Barrow modification, i.e. $\Delta\neq 0$, induces an initial deceleration in the expansion, while changing the asymptotic behaviour, progressively taking the expansion of the system from a power law to an exponential one. In our simulations for values close to $\Delta=0.04$ or higher, it remains exponential with the difference being an initial deceleration  more predominant while having a sharper transition to the asymptotic state. Very large values of the $\Delta$ parameter can bring the system collapse for the given conditions.
	
\section{Conclusions}	

From the works of Hawking and Bekenstein, black hole thermodynamics opened a window to investigate the equations of the gravitational field in a different way, i.e. using the thermodynamics laws. In this paper, gravity-thermodynamics conjecture is used to derived the Friedmann equation considering Barrow entropy. In Barrow model, it is considered the case that quantum-gravitational effects can change actual horizon area of a black hole. In other words, this model constructs a fractal horizon surface by increasing the black hole area. Taking the ideas of gravity-thermodynamics conjecture, $f(R)$ gravity is considered and the first law of thermodynamics in this gravitational model is analyzed. The main study developed in this work was to join the Barrow entropy and $f(R)$ gravity, and from that, analyze the dynamic evolution of the universe during different periods. Corrections for the dark energy components are imposed due to the two models. Our results show that the scalar factor and the state parameter exhibit the expected behavior both for the individual cases and for the union of the Barrow and $f(R)$ models. The results point to a significant influence of the intricacity parameter $\Delta$. However, restrictions related to its values in this new context will be studied in future investigations.

	
	
\section*{Acknowledgments}

	This work by A. F. S. is partially supported by National Council for Scientific and Technological Develo\-pment - CNPq project No. 313400/2020-2; P. S. E. thanks CAPES for financial support.


\global\long\def\link#1#2{\href{http://eudml.org/#1}{#2}}
 \global\long\def\doi#1#2{\href{http://dx.doi.org/#1}{#2}}
 \global\long\def\arXiv#1#2{\href{http://arxiv.org/abs/#1}{arXiv:#1 [#2]}}
 \global\long\def\arXivOld#1{\href{http://arxiv.org/abs/#1}{arXiv:#1}}


\end{document}